\documentclass[letter]{aa} 

\usepackage{graphicx}
\usepackage{txfonts}
\usepackage{multirow}
\usepackage{amsmath}
\usepackage{comment}

\usepackage{graphicx}
\usepackage{gensymb}
\usepackage{epstopdf}
\usepackage{appendix}
\usepackage{siunitx}
\usepackage{textcomp}

\begin{document} 

    \title{Is NGC\,300 a pure exponential disk galaxy?}

   \subtitle{ }

\titlerunning{NGC\,300 outskirts} 
\authorrunning{Jang et al. 2020} 

   \author{In Sung Jang, \inst{1}
          Roelof S. de Jong, \inst{1}
          I. Minchev, \inst{1}
          Eric F. Bell, \inst{2}
          Antonela Monachesi, \inst{3,4}
          Benne W. Holwerda, \inst{5}
          Jeremy~Bailin, \inst{6}
          Adam Smercina, \inst{2}
          Richard D'Souza, \inst{7}
          }

\institute{Leibniz-Institut f\"{u}r Astrophysik Potsdam (AIP), An der Sternwarte 16, 14482 Potsdam, Germany
   \and Department of Astronomy, University of Michigan, 311 West Hall, 1085 South University Ave., Ann Arbor, MI 48109-1107, USA  
   \and Instituto de Investigaci\'on Multidisciplinar en Ciencia y Tecnolog\'ia, Universidad de La Serena, Ra\'ul Bitr\'an 1305, La Serena, Chile
   \and Departamento de Astronom\'ia, Universidad de La Serena, Av. Juan Cisternas 1200 Norte, La Serena, Chile
   \and Department of Physics and Astronomy, 102 Natural Science Building, University of Louisville, Louisville KY 40292, USA
   \and Department of Physics and Astronomy, University of Alabama, Box 870324, Tuscaloosa, AL 35487-0324, USA
   \and Vatican Observatory, 00120 Vatican City State
   }


 
  \abstract
  {
  NGC\,300 is a low-mass disk galaxy in the Sculptor group. In the literature, it has been identified as a pure exponential disk galaxy, as its luminosity profile could be well fitted with a single exponential law over many disk scale lengths (Type I).
  We investigate the stellar luminosity distribution of NGC\,300 using {\em Hubble Space Telescope} (HST) archive data, reaching farther and deeper than any other previous studies.
  Color magnitude diagrams show a significant population of old red giant branch (RGB) stars in all fields out to $R\sim19$~kpc ($32\arcmin$), as well as younger populations in the inner regions.
  We construct the density profiles of the young, intermediate-aged, and old stellar populations.
  We find two clear breaks in the density profiles of the old RGB and intermediate-aged stars: one down-bending (Type II) at $R\sim5.9$ kpc, and another up-bending (Type III) at $R\sim8.3$ kpc.
  Moreover, the old RGB stars exhibit a negative radial color gradient with an up-bending at $R\sim8$~kpc, beyond which the stellar populations are uniformly old ($>$7~Gyr) and metal-poor ($\rm[Fe/H] = -1.6^{+0.2}_{-0.4}$ dex).
  The outer stellar component at $R\gtrapprox8$ kpc is, therefore, well separated from the inner disk in terms of the stellar density and stellar populations.
  While our results cast doubt on the currently established wisdom that NGC\,300 is a pure exponential disk galaxy, a more detailed 
  survey should be carried out to identify the outskirts as either a disk or a stellar halo.
  }


   \maketitle

%

\section{Introduction}
Since the early work of \citet{fre70}, it has been recognized that disk galaxies have an exponential light distribution either without (Type~I) or with a down-bending (Type~II) break.
This classification has more recently been augmented with Type III profiles that have an upward bending in the outer light profile \citep{erw05}.
Various models have been proposed to account for the observed features.
The exponential nature of the disk profile has often been ascribed  to the 
specific angular momentum of the protogalactic cloud \citep{van87}, 
star formation with viscous evolution \citep{fer01}, and
radial stellar scattering \citep{elm16}.
Models for the disk breaks considered 
the angular  momentum  limit \citep{van87},
star  formation threshold \citep{ken89}, 
disk resonances \citep{min12}, and 
minor mergers 
 \citep{you07}.
The very outskirts of disks form sensitive test areas for these models, 
because their evolution is 
more extreme 
due to the low disk self-gravity and the lack of gas for star formation.

Most exponential disk formation models above have difficulty explaining very extended stellar disks, because gas densities are too low for star formation and any (dark matter subhalo) interactions would quickly destroy these tenuous disk outskirts. 
However, observations have shown that there is a population of galaxies that have an exponential light profile traceable out to $\sim$10 disk scale-lengths with no sign of a break or a separate halo component: 
NGC\,5833 - \citet{bar97}, NGC\,4123 - \citet{wei01}, and NGC\,300 - \citet{bla05}. 
Measuring the galaxy light profile in the faint outer regions is very difficult, but taken at face value, such a population of pure exponential disk galaxies would present a major challenge to the known disk formation models. 

In this letter we focus on NGC\,300 
with an up to now known pure exponential light profile \citep{bla05, vla09}.
We present the density profile of NGC\,300 
and report that 
the galaxy has two measurable breaks.

\begin{figure}
\centering
\includegraphics[scale=0.8]{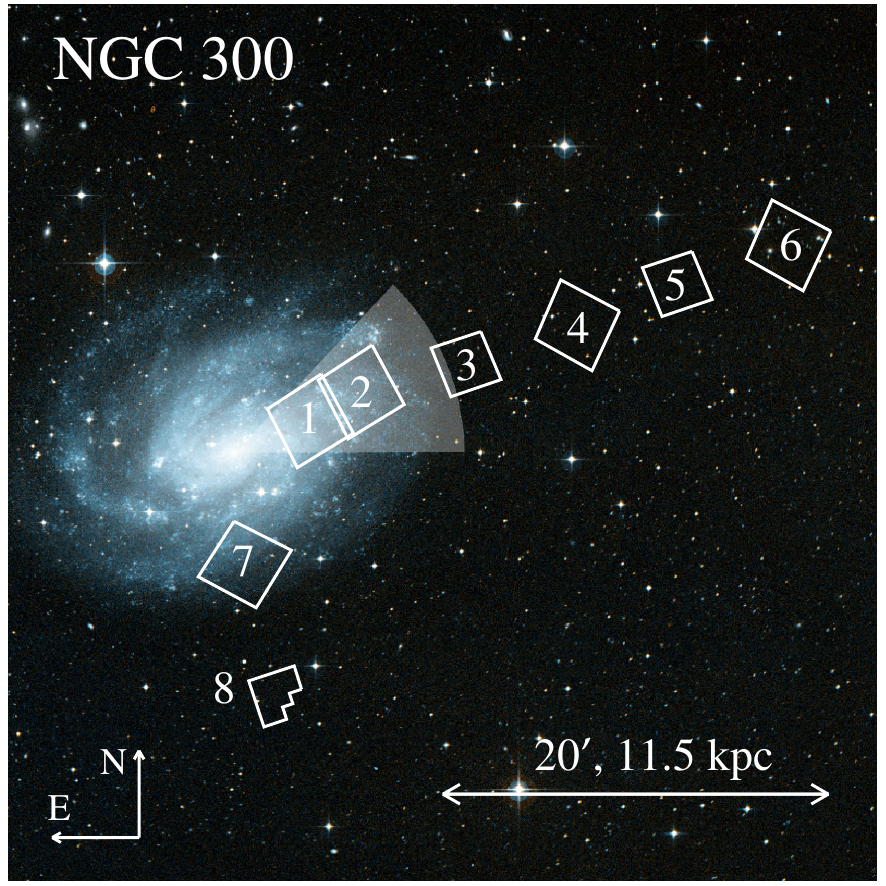}
\caption{
Finding chart for NGC\,300.
The eight {\em HST} fields used in this study are marked on the Digitized Sky Survey image. 
The shaded area indicates the region we used to derive the integrated light profile from the Spitzer $3.6\mu m$ data.
}
\label{fig1}
\end{figure}

\begin{figure*}
\centering 
\includegraphics[scale=0.27]{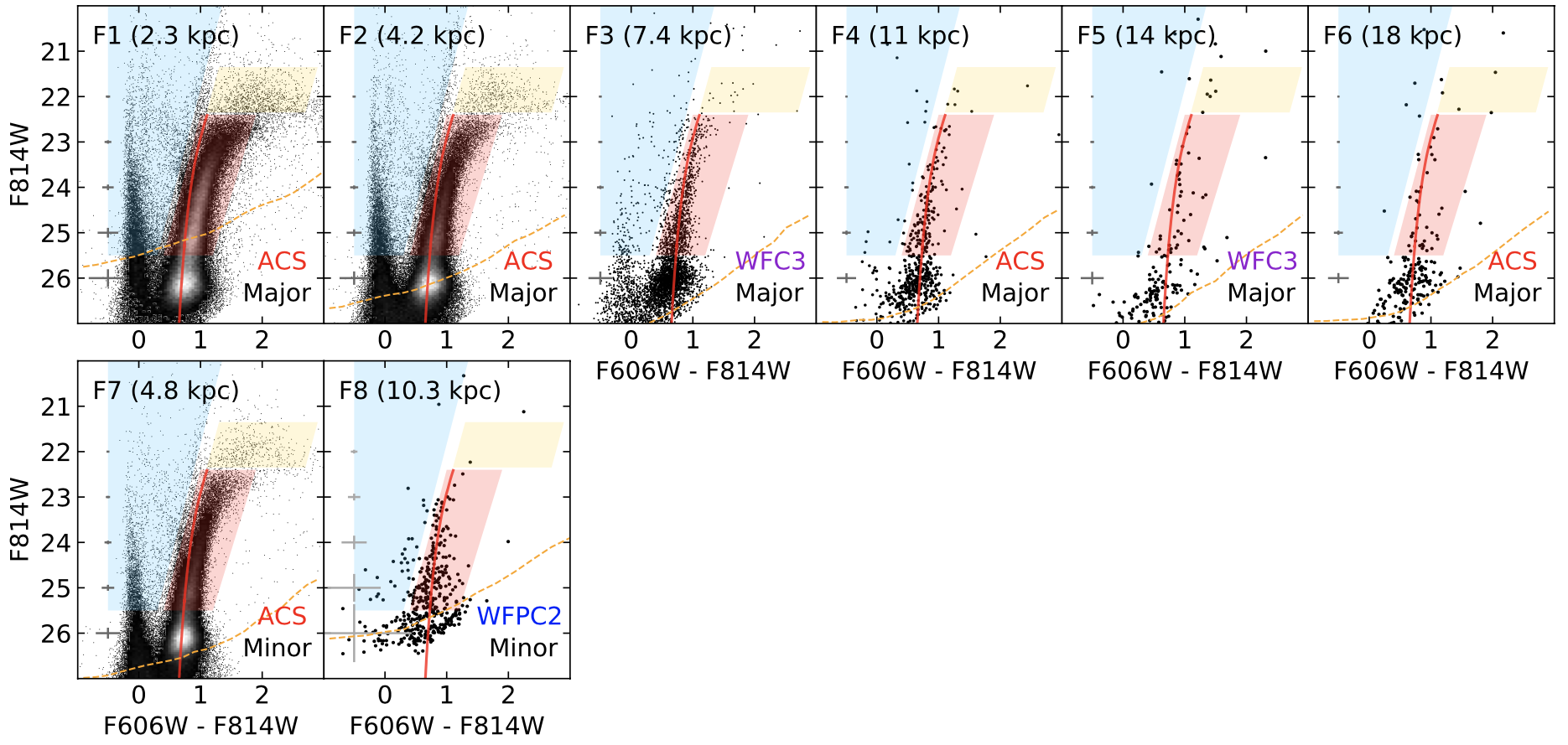}

\caption{
CMDs of the {\em HST} fields along the major (top) and minor (bottom) axes of NGC\,300. 
The field IDs and projected galactocentric distances are marked in each panel.
The projected distances for the minor axis fields 
are corrected for the disk inclination ($i=45\degree$).
Blue, yellow, and pink shaded regions  
represent selection bins for the young (MS and HeB), intermediate-aged (AGB), and old (RGB) stellar populations, respectively.
A stellar isochrone for 10~Gyr age with $\rm [Fe/H]=-1.6$ dex (solid line) and the 50\% completeness limit (dashed line) are overlaid in each CMD. 
}
\label{fig2}
\end{figure*}

\begin{figure*}
\centering
\includegraphics[scale=0.70]{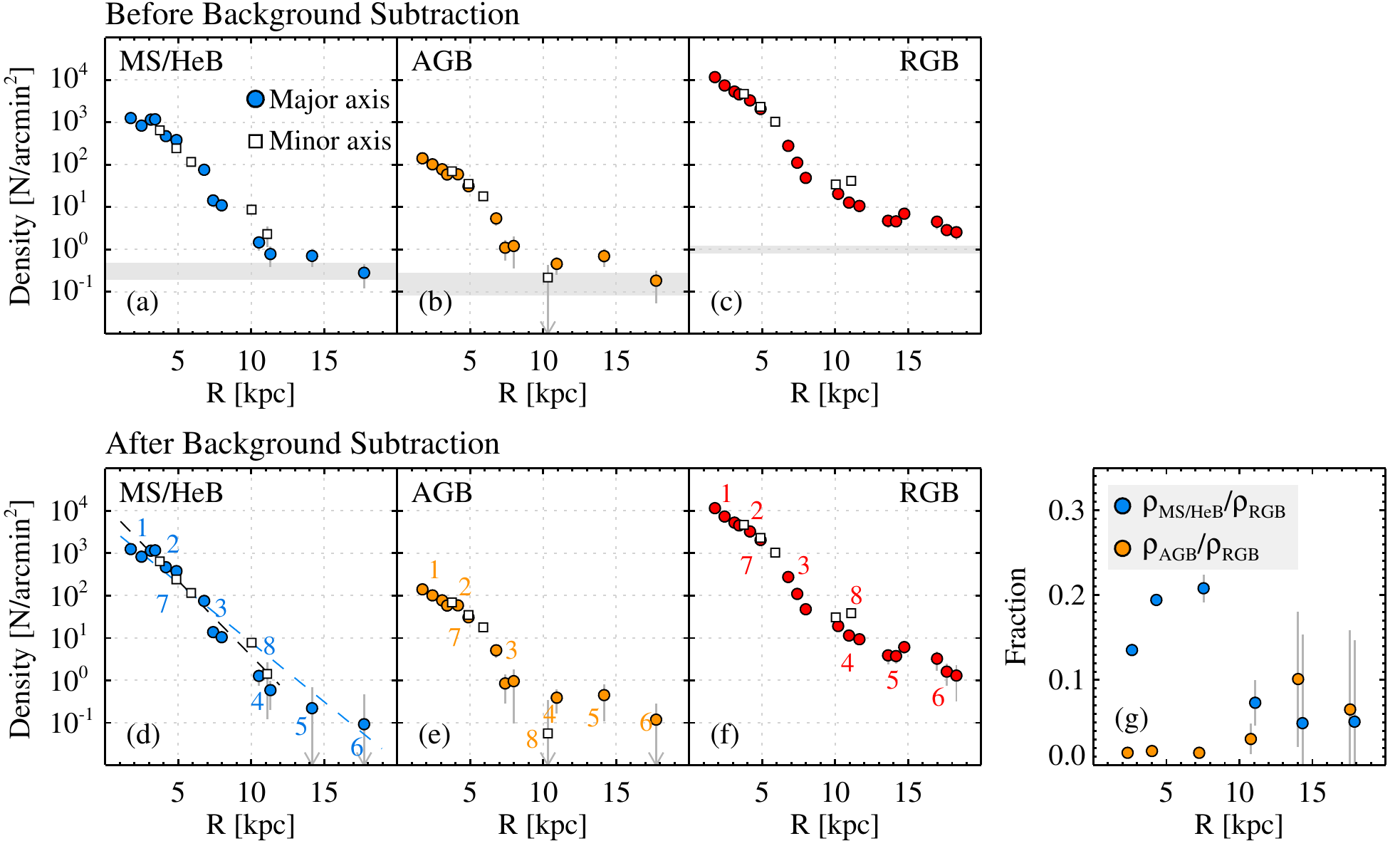}
\caption{(Top) Radial star count profiles for the MS/HeB (a), AGB (b), and RGB (c) stars in NGC\,300.
Circles and squares indicate the profiles along the major and minor axis of the galaxy, respectively. 
The approximate background levels of each population  
are marked by shaded regions. 
(Bottom) Same as top, but after background subtraction. Numbers in each panel indicate field IDs.  
Dashed lines in panel (d) represent the exponential fit of the major and minor axes profiles. 
(g) Relative densities of MS/HeB and AGB stars with respect to RGB stars along the major axis. 
}
\label{fig3}
\end{figure*}

\section{Data and Data Reduction}

We used images of eight {\em HST} fields around NGC\,300 available in the archive. 
Images were taken with three instruments (ACS/WFC, WFC3/UVIS, and WFPC2)  
in $F606W$ and $F814W$ filters.  
Locations of the {\em HST} fields  
are shown in {\bf Figure~\ref{fig1}} and information about the observations can be found in {\bf Table~\ref{taba1}}. 

We obtained photometry of resolved stellar objects 
using the GHOSTS pipeline, which is based on DOLPHOT \citep{dol00}.
After obtaining the raw DOLPHOT output catalogs, we applied photometric cuts to select point sources. 
Details of the pipeline and data reduction procedures can be found in our previous papers \citep{rad11, mon16, jan20}. 
Extensive artificial star tests were made to estimate uncertainties and recovery rates of our photometry.  
We converted the WFC3 and WFPC2 photometric systems to the ACS system, using equations in \citet{jan15} and \citet{sir05}, respectively.

\section{Results}
\subsection{Color magnitude diagrams (CMDs) of resolved stars}
 
In {\bf Figure~\ref{fig2}} we display CMDs of stars in the eight {\em HST} fields.  
Shaded regions indicate the CMD bins we defined to select distinct stellar populations: main sequence (MS) and upper helium-burning (HeB) stars, typically younger than 300 Myr (blue region), bright asymptotic giant branch (AGB) stars with ages of $1-3$~Gyr (yellow region), and red giant branch (RGB) stars mostly older than 3 Gyr (pink region). The CMDs show a gradual change in stellar populations from inner to outer regions as expected. Fields F1, F2 and F7 in the inner regions contain a large population of young MS  
and red HeB stars. 
These features are also visible in F3 at $R=7.4$ kpc, but are much weaker in the fields beyond $R\approx10$ kpc.

We detect a significant population of old RGB stars in all the survey fields, indicating that they extend out to at least 18~kpc. The RGB stars show a radial color gradient, such that the mean RGB color becomes bluer with increasing radial distance. The narrow sequence of RGB stars in the fields beyond 10~kpc (F4, F5, F6, and F8) matches well with a stellar isochrone of 10~Gyr age with $\rm[Fe/H] = -1.6$ dex in the Padova models \citep{bre12} (red line).

The CMDs of NGC\,300 exhibit a dense clump of stars at 
$\rm F814W\sim26.2$~mag, which is the red clump (RC). 
The weak clustering of stars at $\rm F814W\sim25.2$~mag, seen in Fields F5 and F6, is likely the RGB bump. We used these features to estimate the mean stellar age of the galaxy outskirts in Section 3.3.

\subsection{Radial number density profiles of resolved stars}

In {\bf Figure~\ref{fig3}}, we present the radial star count profiles for the three stellar populations (MS/HeB, AGB, and RGB) along the major (circles) and minor (squares) axes of NGC\,300, corrected for the photometric incompleteness using the artificial star data.
For the fields that have enough stars ($N\gtrsim15$), we subdivided them into two or three regions to get a better spatial sampling of the radial profile.
We measured background levels (shaded regions) using empty fields and provide technical details in Appendix A.
Briefly, we used a number of empty fields that are located at a high galactic latitude, similar to the case of the NGC\,300 fields.
These empty fields are expected to be dominated by foreground stars or unresolved background galaxies.
The background level and its uncertainty were taken to be the mean and standard deviation of the source densities in the individual empty fields, respectively.

In the bottom-left panels of {\bf Figure~\ref{fig3}}, we plot the background-subtracted radial star counts.  
The error bars are from Poisson noise of star counts and the background estimation errors.
We trace the young stellar populations out to at least 12~kpc along the major axis (left panel).  
We fitted the major and minor axis profiles with an exponential law (dashed lines), resulting in scale-lengths for the young stellar component of $h_{d,major} = 1.4\pm0.1$ kpc and $h_{d,minor} = 1.2\pm0.1$ kpc. 
While the fit is good along the radially-limited minor axis, an exponential model is clearly a poor fit along the major axis.  

Both the AGB and RGB star count profiles in panels (e) and (f) show 
two clear breaks along the major axis: down-bending, or Type II break, at $R \sim 6$~kpc and up-bending, or Type III break, at $R \sim 8$~kpc. 
The outermost component beyond the Type III break is discussed in Section 4.2.  
Turning our attention towards the minor axis, we find that the sign of break is not clear, mainly due to the short radial coverage with a sparse sampling. 

The bottom-right panel of the figure shows the relative number of MS/HeB stars with respect to RGB stars. 
The ratio of MS stars becomes very low beyond the Type~III break radius, indicating that the outskirts of NGC 300 host substantially less star formation (in a relative sense) than its inner parts.  
Intriguingly, the ratio of AGB stars to RGB stars tentatively appears to be larger towards NGC 300’s largest major axis radii, although the measurement errors are large.

\begin{figure*}
\centering 
\includegraphics[scale=0.80]{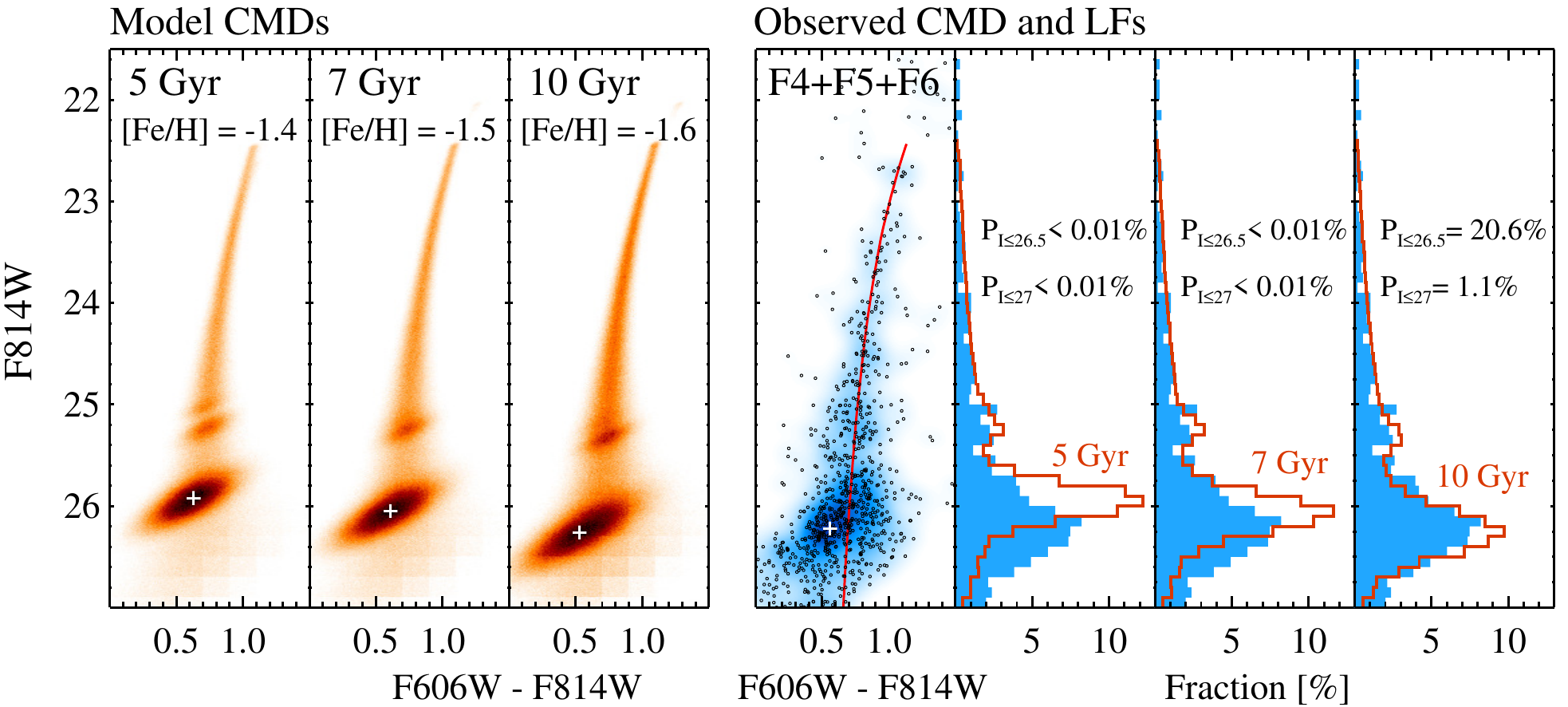}

\caption{(Left three panels) Hess diagrams for the simulated 5, 7, and 10 Gyr simple stellar populations at the distance of NGC\,300.  
The highest density regions of the RC are marked by crosses. Note the gradual change of the RC magnitudes as a function of age.
(Right four panels) Observed CMD of F4, F5 and F6 and its LF (filled histograms) overplotted with the model LFs (open histograms).
The red curved line in the CMD represents the stellar isochrone for age = 10 Gyr and $\rm[Fe/H] = -1.6$ in the Padova models.
The probability (P) of the Kolmogorov-Smirnov test for the stars brighter than F814W = 26.5 and 27 mag are marked in the panels with LFs. 
}
\label{fig4}
\end{figure*}

\subsection{Mean age of the NGC\,300 outskirts}

The absolute magnitude of the RC is known to depend on the age and metallicity of stellar populations \citep{gir01}.  
Three of the eight {\em HST} fields around NGC\,300 (F4, F5, and F6) show an RC above the 50\% completeness level with a dominant old RGB population.  
These have approximately the same color at the Tip of the RGB (TRGB). 
We generated three groups of artificial stars with ages of 5, 7, and 10~Gyr, using the Padova models. Each group contains about two million artificial stars that follow the stellar initial mass function of \citet{cha03}. 
These were shifted according to the distance modulus $(m-M)_0=26.52\pm0.05$ \citep{dal09} and the foreground reddening $E(B-V)=0.011$ \citep{sch11}
of NGC\,300, as well as convolved with photometric uncertainties and incompleteness, similarly to the real stars in the {\em HST} fields.

The left three panels of {\bf Figure \ref{fig4}} show the density-coded CMDs (Hess diagrams) of the artificial stars. 
We adopted metallicities of [Fe/H] = --1.4, --1.5 and --1.6 dex, 
as marked in the figure, 
so that all the model CMDs have approximately the same color at the TRGB of 
$(F606W - F814W)_{\rm TRGB} = 1.1$. 
This color is also consistent with those of the observed CMDs.
All the model CMDs show a prominent RC. The location of the RC changes gradually with age: $F814W \sim$ 25.9, $\sim$ 26.1, and $\sim$ 26.3~mag for the 5-, 7-, and 10-Gyr populations, respectively. There are also two small over-densities above the RC in the 5-Gyr model, which are identified by the AGB bump (at $F814W\sim25.0$ mag) and the RGB bump (at $F814W=25.2$~mag). These two bumps almost overlap in the 7~Gyr and 10-Gyr models at $F814W\sim 25.2$~mag.

On the right side of {\bf Figure \ref{fig4}}, we display a composite CMD of the three observed outer fields (F4, F5, and F6), and its luminosity function (LF) in comparison with the LFs of the model CMDs.
The observed LF shows the best match with the 10 Gyr model LF.
A Kolmogorov-Smirnov test confirms that the oldest 10 Gyr model is preferred than the other younger models, regardless of the faint boundary of the input stars (F814W = 26.5 and 27.0 mag), indicating that the NGC\,300 outskirts is older than 7 Gyr and probably as old as $\sim$10 Gyr.

\subsection{Radial distributions of the RGB stars}
 
\begin{figure}
\centering
\includegraphics[scale=0.76]{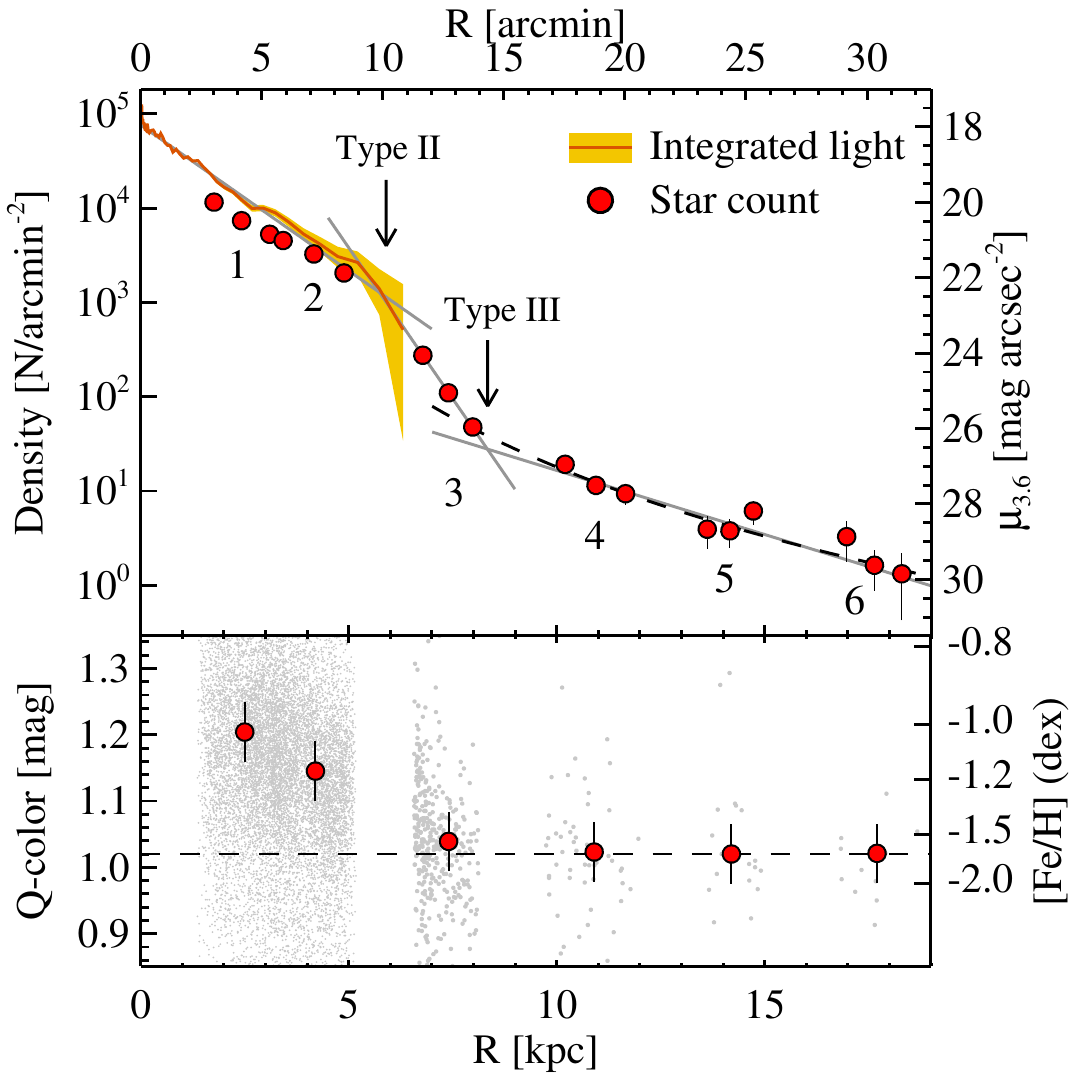}
\caption{(Top) 
Stellar density profile along the NGC\,300 major axis.
RGB star counts are shown by circles. 
The orange line with a shaded region indicates the integrated light profile in the Spitzer $3.6\mu m$ band.  
Surface brightness is shown on the right y-axis, which can be converted to $V$-mag by adding $\sim$2.0~mag. 
Solid lines indicate fits to the resulting three regions using exponential-law models. 
Also shown is a fit to the outermost component with a power-law (dashed line).  
(Bottom) The $Q$-color distribution of the bright RGB stars with $F814W\leq 24.5$~mag. 
Individual stars and their median colors are indicated by gray dots and red circles, respectively. 
}
\label{fig5}
\end{figure}

In the top panel of {\bf Figure \ref{fig5}} we show the radial density distribution of RGB stars along NGC\,300's major axis.
The profile is background subtracted and also corrected for the photometric incompleteness, where completeness is mostly higher than 80\% except for the inner 5~kpc regions ($\sim$60\%).  
We also plot the integrated light profile we derived from the Spitzer $3.6\mu m$ image, 
using the wedge indicated by the shaded area in Figure~\ref{fig1} ($\phi = 293^\circ \pm 22.5^\circ$), to get a similar spatial sampling with the RGB counts. The integrated light profile was scaled to the RGB density using the Padova stellar models.  
We generated a well-populated CMD for old (10 Gyr) and low-metallicity ([Fe/H] = --1.6~dex) stars and derived a relation between the $3.6\mu m$ luminosity and the number of RGB stars in the CMD bin. 

The star count and the integrated light profiles overlap in a region between $R = 2$ and 5~kpc. The two independent profiles show a slight offset of $\sim$0.5 mag in the inner region. Such an offset is naturally expected as the younger stellar populations in this radial range will contribute to the integrated light.  
We divided the profile into three components based on the two breaks 
and fitted each component with an exponential-law model. Derived scale-lengths (fitting range) for the inner to outer components are: 
$h_{d,inner}$ =  $1.4\pm0.1$ kpc ($1< R$ [kpc] $\leq 5$), 
$h_{d,middle}$ = $0.7\pm0.1$ kpc ($6< R$ [kpc] $\leq 8$), and 
$h_{d,outer}$ = $3.2\pm0.5$ kpc ($9< R$ [kpc] $\leq 19$).
The fit for the inner component was made using the integrated light only, while that for the middle one is based on both the integrated light and star count data, taking into account their uncertainties. 
The fits appear to be stable; changing the fitting range of each component by $\pm1$ kpc gives statistically consistent results within $1\sigma$ level.
When we fit the inner component using the star count data, we find $h_{d,inner}$ =  $1.9\pm0.1$ kpc.
The break radii are identified at $R \sim 5.9$ kpc (Type II) and $8.3$ kpc (Type III).

We selected the bright RGB stars with $F814W \leq 24.5$~mag and derived the radial color profile, as shown in  the bottom panel of {\bf Figure \ref{fig5}}.
The equivalent color of the RGB at $F814W=23.0$ mag, accounting for the slope of the RGB following \citet{mon13} ($Q$-index),  
was used to get the median color with a lower scatter. 
The color profile shows a gradual decrease out to $R \sim 8$ kpc, where the Type III break was detected in the density profile. 
Beyond the 8 kpc radius, the color profile does not show a measurable gradient and converges to $(F606W - F814W)_0 = 1.02\pm0.04$. 
The error budget includes systematic uncertainty due to photometric calibration.
We can convert this median color to metallicity using the relation by \citet{str14} with $\rm [\alpha/Fe]=0.3$, which results in $\rm[Fe/H] = -1.6^{+0.2}_{-0.4}$ dex.

\section{Discussion}
\subsection{Comparison with previous studies}
Studies of the NGC\,300 outskirts have a long history. 
Early work by \citet{dev62} and \citet{car85} using photographic plates reported that its luminosity profile falls exponentially down to $\mu_B \sim 26$ mag arcsec$^{-2}$. Their profiles, averaged over azimuthal angle, reached out to $R \sim 13\arcmin$, which is still one of the largest radial extents taken from integrated light. 
In addition to the global exponential fall off, however, 
they also noted slight variations in the profiles.
\citet{dev62} stated that \textit{"The steeper slope beyond $r^* \sim 10\arcmin$ may or may not be real"}.
A similar profile steepening is also seen in Fig. 14 of \citet{car85}. 
More recent studies with CCDs presented an integrated light profile of NGC\,300 \citep{kim04, mun07, lai16}. However, those ended at $R = 9\arcmin - 10\arcmin$, making it difficult to discuss the outer disk properties.   

\citet{bla05} used resolved star counts from Gemini/GMOS combined with the integrated light from \citet{car85}, 
finding that NGC\,300's light profile is well described by a single exponential law out to $R \sim 24\arcmin$ (14 kpc).
However, they had images in only one passband ($r\arcmin$) and no other information (e.g., on stellar populations) could be derived. 
They revisited NGC\,300 in \citet{vla09} with a multi-band ($g\arcmin$ and $i\arcmin$) observations using the same instrument;  
it was confirmed again that the stellar density profile is well approximated by a single exponential law. \citet{vla09} also derived the metallicity profile of the RGB stars showing a monotonic decrease to $R\sim15\farcm5$ and a flattening or an upturn thereafter, with a mean metallicity of the outer component of $\rm[Fe/H] \sim -1.0$ dex.

Our result can be contrasted with the above two studies based on ground-based resolved star counts. Our stellar density profile reaches out to $R\sim32\arcmin$ (19 kpc), or about 30\% farther than those previous studies, 
 and shows two measurable breaks.  
 The median metallicity of the outer component we derived is $\rm[Fe/H] = -1.6^{+0.2}_{-0.4}$ dex, 
 which is 
  significantly lower than the value by \citet{vla09}.  
 However, the {\em HST} fields used in this work are placed along the opposite side (north-west) of NGC\,300's major axis, compared to the GMOS fields used in the previous studies (south-east), 
 which makes it difficult to compare the density profiles directly.
The measurable difference in both the density profile and abundance could be due to the asymmetric distribution of the NGC\,300 outskirts or different analysis methods applied in each study (e.g., stellar model, background estimation), but the exact origin is unclear requiring further study.

\citet{hil16} studied the young stellar disk of NGC\,300 using the same {\em HST} data as in our fields F3 -- F6. 
Using a CMD fitting technique, they isolated a stellar population younger than 200 Myr and found an unbroken young disk profile out to $R = 11$~kpc (with an upper limit out to  14 kpc) 
with a scale-length of $h_d = 1.4\pm0.1$ kpc. 
While the global exponential decay of the young density profile is reproduced in this study, the profile appears to be a poor fit with a single exponential law over the whole radial extent (Figure~\ref{fig3}, bottom left).

The Type II break we find at $R \sim10\arcmin$ in the AGB and RGB profiles is also supported by a down-bending in the UV profile of NGC\,300 found by \citet{rou05}  and \citet{gil07}. 
Taken at face value, the youngest stellar disk of the galaxy is also likely truncated.

\subsection{Identification of the outer stellar component: extended disk or stellar halo?}

The outer stellar component beyond the up-bending 
is well separated from the inner ones, in terms of both its stellar density and its stellar populations. The identification of this outer component, as either an extended stellar disk or a stellar halo, is important to understand the growth history of the galaxy outskirts. 
Unfortunately, it is not easy to distinguish between both options based on the limited information we have, especially with the lack of measurements along the minor axis of the galaxy.  

Interpreting the outer component as disk stars, we note that NGC\,300 has an extended $HI$ disk reaching out to at least $R\sim20$~kpc \citep{wes11}, which may be the source of our younger population reaching out to $R\sim17$~kpc. 
Explaining the upturn in the NGC\,300's outer stellar component is then still hard to explain as it consists mainly of old stars, which would be hard to transport to these extreme radii by the radial migration of earlier disk stars alone. \cite{min12} used N-body tree-SPH simulations to show that a combination of Type II + III breaks, as found here, can result from gas accretion in the outer region, which has the effect of strongly increasing the eccentricity of the orbits of the existing stellar population beyond Type III (see their Fig.~17). The anti-truncation can also result from the perturbation of a minor merger, provided the existence of a gas-rich disk \citep{you07}. A successful NGC\,300 pure exponential disk model should be able to explain both the double break in the density profile and the old stars with a flat color/metallicity profile beyond the Type III break.

Alternatively, the outermost component could be a stellar halo. Fitting to it a power-law model (curved dashed line in {\bf Figure~\ref{fig5}}) resulted in a {2D density} slope of $\alpha_{major}= -4.2\pm0.5$, which is in good agreement with $-5.3\leq\alpha_{major} \leq-2.7$, measured for the stellar halos of nearby Milky Way mass disk galaxies \citep{har17}.
Integrating numerically the power-law profile from 10 kpc to 19~kpc, we obtained an observed total stellar mass of $M_{halo,10-19} = 3\times 10^7M_\odot$.  
Here we assumed 10 Gyr age with $\rm[Fe/H]=-1.6$ dex.
Combining the halo mass with the total stellar mass of NGC\,300 of $M_{gal} = 2\times 10^9 M_\odot$ \citep{mun15,lai16}, we obtain the stellar halo mass fraction, $M_{halo,10-19}/M_{gal} = 1.5\%$.
Extrapolating this observed halo profile further inward, the total stellar halo mass fraction becomes several times larger.  
Such a massive stellar halo would be very unusual in current low-mass disk galaxy models,  
 as low-mass galaxies are assembled from lower-mass dark matter halos with lower star formation efficiency \citep{pur08,cop13}.

\section{Closing remarks}

The nearby galaxy NGC\,300 has been presented in the literature as a prototypical example of a Type I disk galaxy with a pure exponential light profile over many disk scale lengths. 
Most exponential disk formation models have difficultly explaining such an extended stellar disk; accordingly 
we investigated NGC\,300's light distribution  
using high-resolution $HST$ data in the archive.

We constructed density profiles of different stellar populations reaching out to $\sim$19~kpc from the galaxy center. 
We found that the profiles for the AGB and RGB populations show two clear breaks: one down-bending (Type~II) at $R\sim5.9$ kpc, and another up-bending (Type~III) at $R\sim8.3$ kpc. 
The stellar populations beyond the Type~III break radius become predominantly old ($\gtrsim$9 Gyr) and metal-poor ([Fe/H]=-1.6), well separated from the inner disk regions.
 
While our results cast doubt on the currently established wisdom that NGC\,300 is a pure exponential disk galaxy, a more detailed study should be carried out. 
It is expected that wide-field imaging with ground based large telescopes would be useful to map out the global structure of the extended stellar component, providing evidence to the formation history of NGC\,300.

\begin{acknowledgements}
We are thankful to the anonymous referee
for his/her useful comments.
I.S.J. and R.dJ. gratefully acknowledge support from the Deutsches Zentrum f\"ur Luft- und Raumfahrt (DLR) through grant 50OR1815. 
A.S. acknowledges support for this work by the National Science Foundation Graduate Research Fellowship Program under grant No. DGE 1256260. Any opinions, findings, and conclusions or recommendations expressed in this material are those of the author(s) and do not necessarily reflect the views of the National Science Foundation.

This paper is based on observations made with the NASA/ESA Hubble Space Telescope, obtained from the data archive at the Space Telescope Science Institute. STScI is operated by the Association of Universities for Research in Astronomy, Inc. under NASA contract NAS 5-26555. 
I.S.J. is grateful to Florian Niederhofer for his help in Python programming.
AM acknowledges financial support from FONDECYT Regular 1181797 and funding from the Max Planck Society through a Partner Group grant.

This research has made use of the NASA/IPAC Extragalactic Database (NED), which is funded by the National Aeronautics and Space Administration and operated by the California Institute of Technology.
\end{acknowledgements}

%
%

\appendix{}
\section{Contamination estimation} 

NGC\,300 is located at a high galactic latitude of $b \sim 79.4\degree$ in the sky, and its CMDs shown in Figure~\ref{fig2} exhibit a well defined sequence of 
old RGB stars, indicating that the detected point sources are mostly the NGC\,300 stars. 
Nevertheless, there could be a small contribution of contaminating sources that passed the GHOSTS culls, such as foreground stars in the Milky Way and unresolved background galaxies. 
These sources should be properly taken into account to derive the intrinsic radial number density profile of the NGC\,300 stars.
We describe here the method we used for measuring the contamination levels of the NGC\,300 data.

We selected eight fields in empty sky regions taken with each $HST$ instrument used in this study (ACS/WFC, WFC3/UVIS, and WFPC2), as listed in Table \ref{taba1}. 
These fields are located at high galactic latitude with $|b| \gtrsim 40\degree$, and also well away from known stellar objects in the sky (e.g., bright stars, star clusters, tidal streams in the Milky Way halo, bright galaxies, and galaxy clusters).
The fields for ACS/WFC and WFC3/UVIS have also been used for the background control of M101  
 in our previous study \citep{jan20}.
 Table \ref{taba1} also lists the NGC\,300 fields used in this study for reference. 
 
\begin{figure*}
\centering
\includegraphics[scale=0.9]{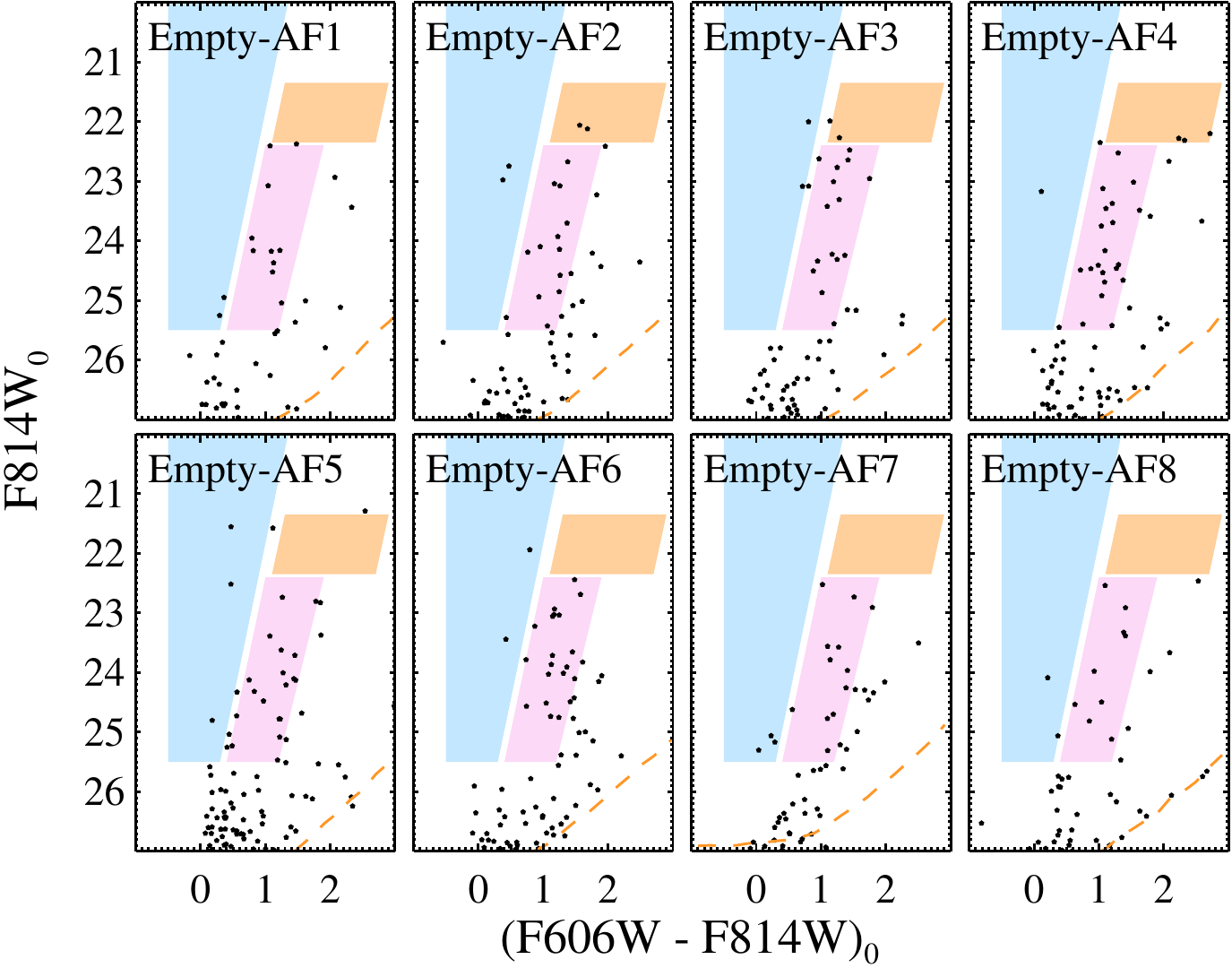}
\caption{{
CMDs for the selected point sources in the eight empty fields taken with ACS/WFC.
We correct for the foreground extinction toward each field using the values in \citet{sch11}.
Shaded regions are the same as those in Figure \ref{fig2}. 
Dashed lines represent the 50\% completeness limits.
}
}
\label{figa1}
\end{figure*}

\begin{figure*}
\centering
\includegraphics[scale=0.9]{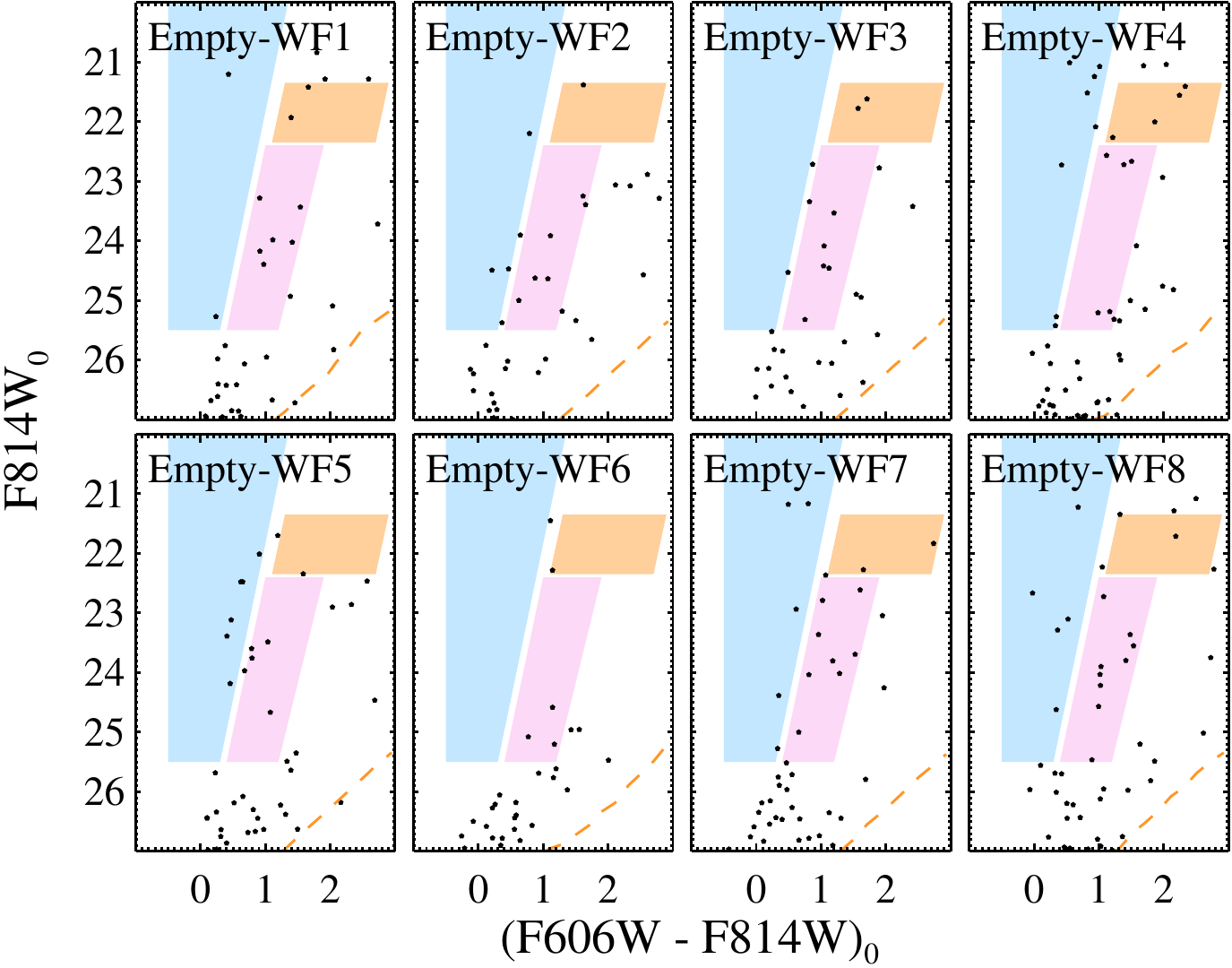}
\caption{Same as Figure  \ref{figa1}, but for WFC3/UVIS fields.}
\label{figa2}
\end{figure*}

\begin{figure*}
\centering
\includegraphics[scale=0.9]{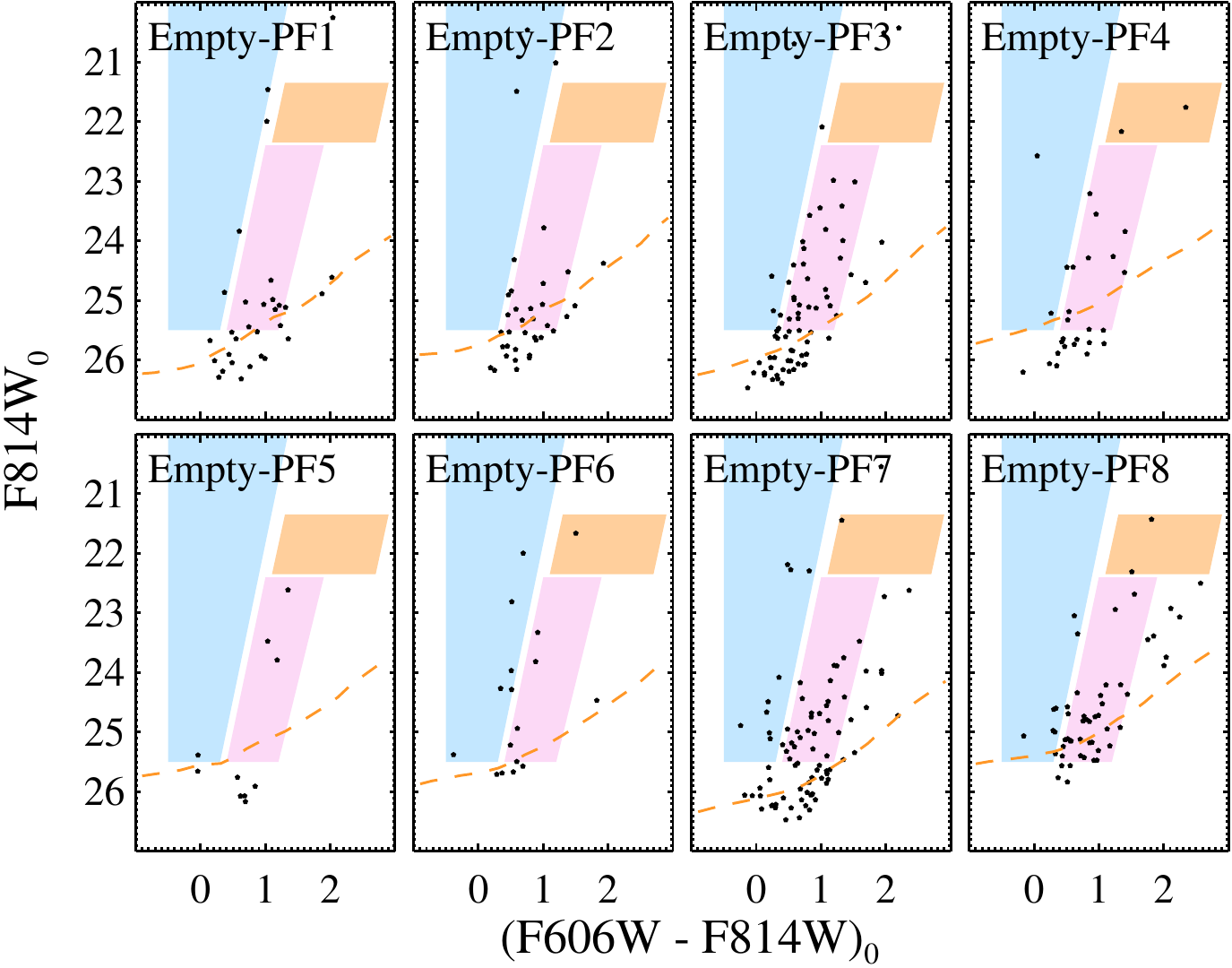}
\caption{Same as Figure  \ref{figa1}, but for WFPC2 fields.}
\label{figa3}
\end{figure*}

We treated these empty sky fields identically to the NGC\,300 fields, deriving point source photometry using the GHOSTS pipeline.
Figures \ref{figa1}, \ref{figa2}, and \ref{figa3} show CMDs of the emtpy fields taken with ACS/WFC, WFC3/UVIS, and WFPC2, respectively.
Shaded regions in the figures indicate the population bins (Ms/HeB, AGB, and RGB) used for the NGC\,300 data.
The sources presented in the CMDs are considered to be the contaminating sources, either foreground stars or unresolved background galaxies.
We calculated the spatial number densities of these sources considering the photometric incompleteness and the effective field area, as listed in Table \ref{taba2}.
Here the errors are taken to be the standard deviation of the individual densities.
We used these values to correct for the contribution of contaminating sources in the NGC\,300 profiles.

\begin{table*}
\caption{Summary of $HST$ data of the control fields and NGC\,300 used in this study}
\centering
\begin{tabular}{lrrrrrrr}
\hline
\hline
Field & R.A. & Decl & $b$ & Inst. & \multicolumn{2}{c}{Exposure Time(s)} & Prop ID  \\
 & (2000.0) & (2000.0) & ($^{\circ}$) & & $F606W$  & $F814W$ \\
\hline

Empty-AF1 & 01 05 33.92 &--27 37 03.0 &--86.8& ACS	& 4,080 & 4,080 & 9498 \\ 
Empty-AF2 & 14 17 02.01 &  52 25 02.7 &  60.1& ACS	& 2,260 & 2,100 & 10134 \\ 
Empty-AF3 & 14 19 18.17 &  52 49 24.6 &  59.6& ACS	& 2,260 & 2,100 & 10134 \\ 
Empty-AF4 & 14 21 35.89 &  53 13 48.3 &  59.0& ACS	& 2,260 & 2,100 & 10134 \\ 
Empty-AF5 & 14 25 15.26 &  35 38 20.5 &  68.3& ACS	& 4,500 & 4,500 & 10195 \\ 
Empty-AF6 & 10 56 31.88 &--03 41 18.2 &  48.5& ACS	& 2,160 & 2,160 & 10196 \\ 
Empty-AF7 & 12 38 18.65 &  62 20 51.0 &  54.7& ACS	& 2,092 & 2,092 &  13779 \\ 
Empty-AF8 & 03 32 43.58 &--27 47 55.9 &--54.4& ACS	& 2,556 & 2,556 & 11563 \\   

\hline 
Empty-WF1 &  4 41 48.87 &--38 05 08.5 & --41.0 & WFC3	& 2,500 & 2,500 & 13352 \\  
Empty-WF2 &  8 53 59.34 &  43 52 07.2 &   40.0 & WFC3	& 4,400 & 4,400 & 13352 \\  
Empty-WF3 &  1 25 36.40 &--00 00 47.5 & --61.7 & WFC3	& 3,200 & 6,400 & 13352 \\  
Empty-WF4 & 15 00 22.92 &  41 28 03.9 &   60.0 & WFC3	& 2,650 & 2,000 & 14178 \\  
Empty-WF5 & 10 51 21.54 &  20 21 27.1 &   61.8 & WFC3	& 4,050 & 3,950 & 14178 \\  
Empty-WF6 & 12 08 35.15 &  45 44 20.5 &   69.5 & WFC3	& 3,230 & 3,180 & 14178 \\  
Empty-WF7&  3 33 23.16 &--40 57 27.8 & --54.1 & WFC3	& 3,150 & 3,150 & 14178 \\  
Empty-WF8&  9 47 02.82 &  51 26 12.5 &   47.8 & WFC3	& 2,750 & 2,400 & 14178  \\ 

\hline

Empty-PF1 &  12 36 00.75 & 62 10 23.3 & 54.9 & WFPC2	& 2,200 & 2,200 & 7410 \\  
Empty-PF2 &  03 32 28.78 &--27 45 57.7 & --54.4 & WFPC2	& 2,000 & 2,000 & 8809 \\  
Empty-PF3 &  14 17 37.00 & 52 27 24.7 & 60.0 & WFPC2	& 2,100 & 2,200 & 5090 \\  
Empty-PF4 &  14 31 40.35 & 36 18 00.5 & 66.9 & WFPC2	& 2,000 & 2,000 & 11513 \\ 
Empty-PF5 &  01 29 36.76 & --16 05 45.3 & --75.8 & WFPC2	& 2,000 & 2,000 & 9676 \\ 
Empty-PF6 &  13 59 18.11 & 62 34 55.9 & 52.8 & WFPC2	& 2,200 & 2,000 & 9676 \\  
Empty-PF7 &  14 35 33.58 & 25 18 08.9 & 66.6 & WFPC2	& 2,400 & 2,000 & 5370 \\  
Empty-PF8 &  14 45 10.85 & 10 02 55.8 & 58.2 & WFPC2	& 1,700 & 1,700 & 5091 \\  

\hline

NGC\,300-F1	& 00 54 32.99 & --37 40 12.9 & --79.4 & ACS	& 1,515 & 1,542 & 10915 \\ 
NGC\,300-F2	& 00 54 21.66 & --37 37 56.3 & --79.5 & ACS	& 1,515 & 1,542 & 10915 \\ %
NGC\,300-F3  & 00 53 53.89 & --37 36 29.6 & --79.5 & WFC3	& 1,000 & 1,500 & 13461 \\ %
NGC\,300-F4  & 00 53 25.03 & --37 34 30.1 & --79.5 & ACS	&   894 & 1,354 & 13461 \\ %
NGC\,300-F5  & 00 52 58.65 & --37 32 20.5 & --79.6 & WFC3	& 1,000 & 1,500 & 13461 \\   %
NGC\,300-F6  & 00 52 29.81 & --37 30 21.0 & --79.6 & ACS	&   894 & 1,354 & 13461 \\ %
NGC\,300-F7  & 00 54 52.10 & --37 46 49.5 & --79.3 & ACS	& 2,400 & 2,548 & 13515 \\ %
NGC\,300-F8  & 00 54 44.17 & --37 53 42.3 & --79.2 & WFPC2	&2,000 & 2,000 & 9086 \\   %

\hline
\end{tabular}
\label{taba1}
\end{table*}

\begin{table*}
\caption{Estimated contamination levels for distinct stellar populations measured from the control fields}
\centering
\begin{tabular}{lrccc}
\hline
\hline
Instrument & MS/HeB & AGB & RGB  \\
& [$N$ arcmin$^{-2}$] & [$N$ arcmin$^{-2}$] & [$N$ arcmin$^{-2}$] \\
\hline
ACS/WFC  &   $0.19\pm0.06$ & $0.06\pm0.10$ & $1.24\pm0.41$ \\
WFC3/UVIS&   $0.48\pm0.34$ & $0.24\pm0.15$ & $0.83\pm0.27$ \\
WFPC2    & $0.88\pm0.61$ & $0.16\pm0.19$ & $3.01\pm2.23$ \\
\hline
\hline
\end{tabular}
\label{taba2}
\end{table*}

\end{document}